\documentclass[aps, prl, twocolumn, superscriptaddress, showpacs]{revtex4}
\usepackage{mathrsfs, amsfonts, amsmath, amssymb, graphicx, bm, epsfig}
\usepackage[colorlinks, linkcolor=blue, anchorcolor=blue, citecolor=red]{hyperref}

\begin{document}
\title{Spin Pumping and Spin-Transfer Torques in Antiferromagnets}

\author{Ran Cheng}
\email{rancheng@utexas.edu}
\affiliation{Department of Physics, University of Texas at Austin, Austin, Texas 78712, USA}

\author{Jiang Xiao}
\email{xiaojiang@fudan.edu.cn}
\thanks{The first two authors contributed equally to this work.}
\affiliation{Department of Physics and State Key Laboratory of Surface Physics, Fudan University, Shanghai 200433, China}

\author{Qian Niu}
\affiliation{Department of Physics, University of Texas at Austin, Austin, Texas 78712, USA}
\affiliation{International Center for Quantum Materials, and Collaborative Innovation Center of Quantum Matter, School of Physics, Peking University, Beijing 100871, China}

\author{Arne Brataas}
\affiliation{Department of Physics, Norwegian University of Science and Technology, NO-7491 Trondheim, Norway}


\begin{abstract}
Spin pumping and spin-transfer torques are two reciprocal phenomena widely studied in ferromagnetic materials. However, pumping from antiferromagnets and its relation to current-induced torques have not been explored. By calculating how electrons scatter off a normal metal-antiferromagnetic interface, we derive pumped spin and staggered spin currents in terms of the staggered field, the magnetization, and their rates of change. For both compensated and uncompensated interfaces, spin pumping is of a similar magnitude as in ferromagnets with a direction controlled by the polarization of the driving microwave. The pumped currents are connected to current-induced torques via Onsager reciprocity relations.
\end{abstract}

\maketitle

A major task of spintronics is understanding the mutual control of spin transport and magnetic properties. This inspires intense studies in fundamental physics which opens new avenues in, e.g., magnetic recording technologies. A new direction in this field aims at harnessing spin dynamics in materials with a vanishing magnetization, such as antiferromagnets (AFs) with compensated magnetic moments on an atomic scale. As compared to ferromagnets (Fs), AFs operate at a much higher frequency in the Tera Hertz (THz) ranges~\cite{ref:Rasing,ref:NiOswitch,ref:Nowak} which makes it possible to perform ultra fast information processing and communication. At the same time, since there are no stray fields in AFs, they are more robust against magnetic perturbations, an attractive feature of AFs for use in next-generation data storage material. However, to build a viable magnetic device using AF, it is vital to find observable effects induced by the rotation of the order parameter. The recent discovery of tunneling anisotropic magnetoresistance in AF may potentially fulfill this demand~\cite{ref:TAMR1,ref:TAMR2}. Nevertheless, in such experiments, the AF is dragged passively by an adjacent F, which is rotated by a magnetic field. Will an AF interact directly with (spin) currents without the inclusion of a F or a magnetic field?

Partial answers are available from recent investigations. While the observation of a current-induced change of the exchange bias on a F$|$AF interface indicates spin-transfer torques (STTs) in AFs~\cite{ref:Tsoi,ref:Urazhdin}, theoretical models of STT have been developed in a variety of contexts~\cite{ref:MacD,ref:Xia,ref:Duine,ref:Arne,ref:Ran,ref:Jacob,ref:Gomonay,ref:Manchon}. To achieve a general understanding of spintronics based on AFs, we recall a crucial insight from well-established ferromagnetic spintronics: STT and spin pumping are two reciprocal processes intrinsically connected~\cite{ref:STTReview1,ref:STTReview2,ref:spinpumping}; they are derivable from each other~\cite{ref:Maekawabook}. To the best our knowledge, all existing studies on AF have focused on STT, whereas spin pumping has received no attention because it seems to be naively believed that the vanishing magnetization spoils any spin pumping in an AF.

\begin{figure}[b]
   \centering
   \includegraphics[width=0.8\linewidth]{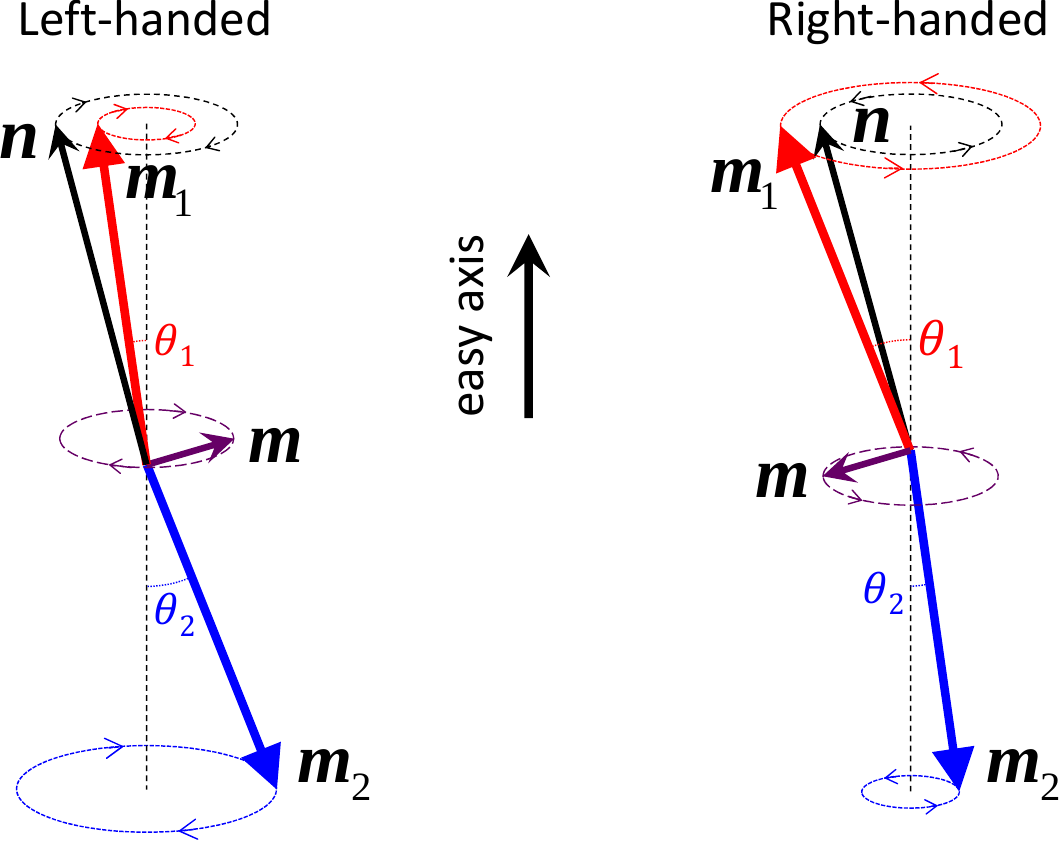}
   \caption{(Color online) The two eigenmodes of Eq.~\eqref{eq:resonance} have opposite chiralities and opposite ratios between the cone angles of $\bm{m}_1$ and $\bm{m}_2$. A magnetic field along the easy axis breaks the degeneracy of the two modes.}{\label{Fig:Modes}}
\end{figure}

Spin pumping is the generation of spin currents by the precessing magnetization~\cite{ref:spinpumping,ref:Maekawabook}. When the magnetization $\bm{m}$ of a F varies in time, a spin current proportional to $\bm{m}\times\dot{\bm{m}}$ is pumped into an adjacent normal (N) metal. In contrast, $\bm{m}$ vanishes in equilibrium in homogeneous AFs and is small even when the system is driven out-of-equilibrium. Instead, it is the staggered field (or N\'{e}el order) $\bm{n}$ that characterizes the system. Does the motion of $\bm{n}$ lead to any pumping effect?

In this Letter, we first argue heuristically that spin pumping from the compensated magnetization of the two sublattices constructively adds up rather than cancel. We confirm this anticipation by exploring electron scattering across a N$|$AF interface, and derive analytically the pumped spin and staggered spin currents. To complete the reciprocal picture, we finally derive the STT due to an applied spin voltage.

\textit{Antiferromagnetic resonance.}--- We consider an AF with two sublattices and an easy axis along $\hat{\bm{z}}$~\cite{ref:Kittel}. The directions of the magnetic moments are denoted by two unit vectors $\bm{m}_1$ and $\bm{m}_2$. The precession of $\bm{m}_1$ and $\bm{m}_2$ are driven by the exchange interaction, the anisotropy, and a magnetic field assumed to be in the $\hat{\bm{z}}$-direction. In units of frequency, they are represented by $\omega_E$, $\omega_A$, and $\omega_H=\gamma H_0$, respectively. The equations of motion are
\begin{subequations}
\label{eq:sublattdamping}
\begin{align}
 \dot{\bm{m}}_1&=\bm{m}_1\times[\omega_E\bm{m}_2-(\omega_A+\omega_H)\hat{\bm{z}}], \\
 \dot{\bm{m}}_2&=\bm{m}_2\times[\omega_E\bm{m}_1+(\omega_A-\omega_H)\hat{\bm{z}}],
\end{align}
\end{subequations}
where additional damping terms will be taken into account only when necessary. In linear response, we decompose $\bm{m}_{1,2}$ into equilibrium and oscillating parts $\bm{m}_1=\hat{\bm{z}}+\bm{m}_{1,\perp}e^{i\omega t}$ and $\bm{m}_2=-\hat{\bm{z}}+\bm{m}_{2,\perp}e^{i\omega t}$, and assume $|\bm{m}_{\perp}|\ll1$. The resonance frequencies are then
\begin{align}
 \omega=\omega_H\pm\omega_R=\omega_H\pm\sqrt{\omega_A(\omega_A+2\omega_E)}, \label{eq:resonance}
\end{align}
and the two corresponding eigenmodes are depicted in Fig.~\ref{Fig:Modes}, which are characterized by different chiralities. From a bird's eye view along $-\hat{\bm{z}}$ of the left-handed (right-handed) mode, both $\bm{m}_1$ and $\bm{m}_2$ undergo a circular clockwise (counterclockwise) precession with $\pi$ phase difference. In the absence of magnetic field, \textit{viz.} $\omega_H=0$, the two modes are degenerate.

\begin{figure}
   \centering
   \includegraphics[width=0.95\linewidth]{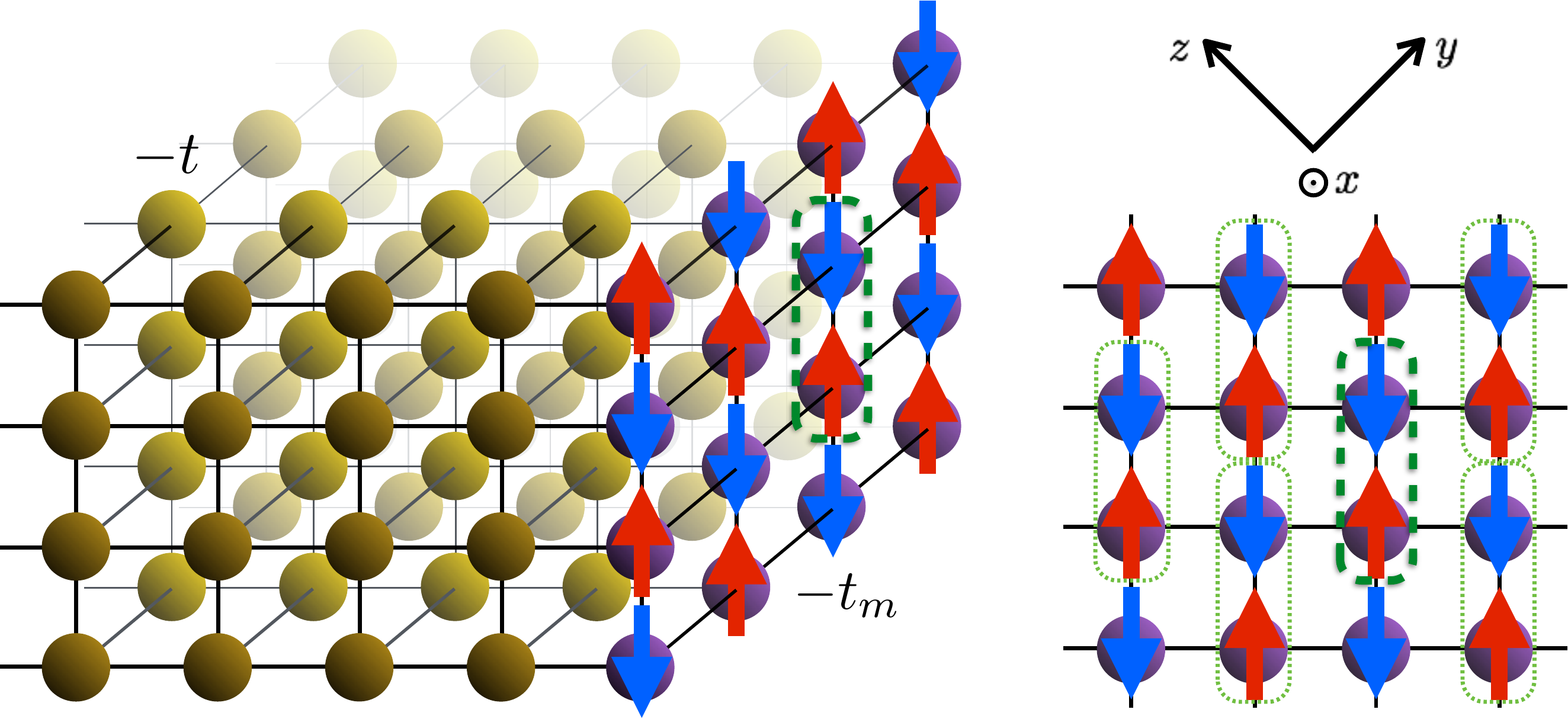}
   \caption{(Color online) A compensated N$|$AF interface with cubic lattice. The interface normal is along $\hat{\bm{x}}$. Unit cells (dotted Green circles) are periodic in the $[0,1,1]$ and $[0,\bar{1},1]$ directions, which are labeled by $\hat{\bm{y}}$ and $\hat{\bm{z}}$, respectively. }{\label{Fig:Interface}}
\end{figure}

A heuristic way to grasp the essential feature of spin pumping by AF is to consider $\bm{m}_1$ and $\bm{m}_2$ as two independent F subsystems. Then spin currents pumped from them will be proportional to $\bm{m}_1\times\dot{\bm{m}}_1$ and $\bm{m}_2\times\dot{\bm{m}}_2$, respectively. From Fig.~\ref{Fig:Modes} we see that $\bm{m}_1\approx-\bm{m}_2$ and $\dot{\bm{m}}_1\approx-\dot{\bm{m}}_2$, thus the contributions from the two are basically the same and add up constructively. As a result, the total spin current is roughly proportional to $\bm{n}\times\dot{\bm{n}}$ where $\bm{n}=(\bm{m}_1-\bm{m}_2)/2$ denotes the staggered field. However, a more careful analysis reveals that the cone angles of $\bm{m}_1$ and $\bm{m}_2$ are different: in the left-handed (right-handed) mode, $\theta_2/\theta_1=\eta$ ($\theta_1/\theta_2=\eta$), where $\eta\approx(1+\sqrt{\omega_A/\omega_E})^2$, so that a small magnetization $\bm{m}$ will be induced, as shown in Fig.~\ref{Fig:Modes}.

Furthermore, scattering channels associated with different sublattices on a N$|$AF interface will mix, thus an AF is not equivalent to two Fs. To what extent the above heuristic picture survives is ultimately determined by the interface scattering of electrons.

\textit{Interface scattering.}--- Typical AF materials are insulators~\cite{ref:MnF2,ref:FeF2} and incident electrons from the normal metal cannot penetrate far. Consequently, only a single atomic layer of AF directly connected to N suffices to describe the dominant contribution to interface scattering. Therefore, the essential physics is captured by modeling the N$|$AF interface as being semi-infinite in the transport direction and infinite in the transverse direction. As illustrated in Fig.~\ref{Fig:Interface}, the interface is compensated, where neighboring magnetic moments are located at different sublattices. The case of an uncompensated interface is analogous to a N$|$F(insulator) interface.

\begin{figure}[t]
   \centering
   \includegraphics[width=0.95\linewidth]{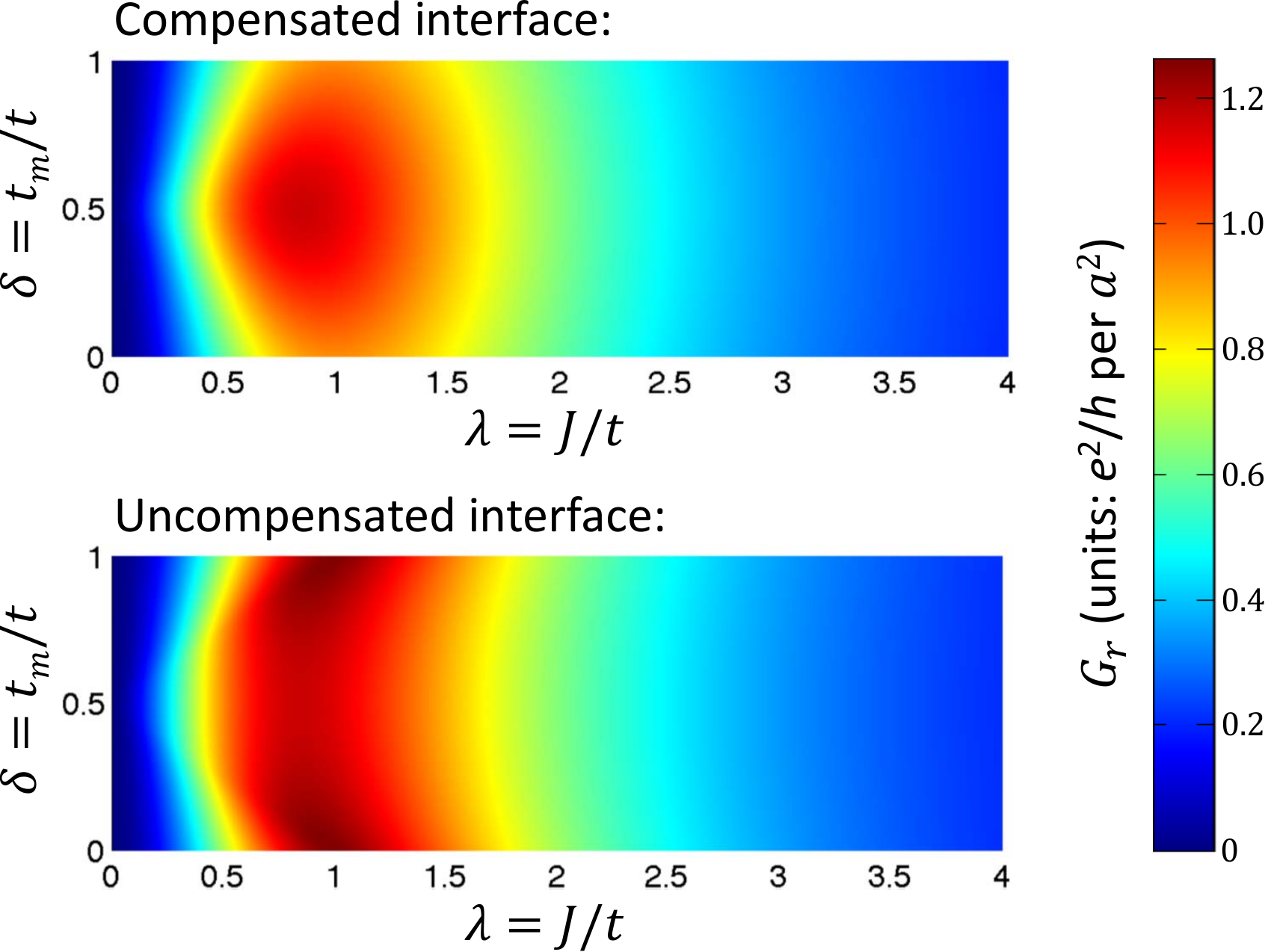}
   \caption{(Color online) Spin mixing conductance $G_r$ as a function of $\lambda$ and $\delta$ in units of $e^2/h$ per $a^2$ for compensated and uncompensated N$|$AF interfaces.}{\label{Fig:DeltaS}}
\end{figure}

Adopting the nearest-neighbor tight-binding model on a cubic lattice, we denote the hopping energy in N and AF by $t$ and $t_m$, respectively. The lattice constant is $a$, and the exchange coupling between conduction electron spins and magnetic moments is $J$, we define the dimensionless energies $\delta=t_m/t$ and $\lambda=J/t$. To linear order in the small $\bm{m}$, the scattering matrix is
\begin{align}
 S=S_0+S_w\hat{\tau}_1\hat{\sigma}_0+\Delta S[\hat{\tau}_3(\bm{n}\cdot\hat{\bm{\sigma}})+\hat{\tau}_0(\bm{m}\cdot\hat{\bm{\sigma}})], \label{eq:Smatrix}
\end{align}
where $\hat{\tau}_{1,2,3}$ are pseudo-spin Pauli matrices for sublattice degree of freedom, $\hat{\bm{\sigma}}$ are spin Pauli matrices, and $\hat{\tau}_0$ and $\hat{\sigma}_0$ are identity matrices. The last two terms of Eq.~\eqref{eq:Smatrix} with a common coefficient $\Delta S$ are spin-dependent and represent Umklapp and normal scatterings, respectively~\cite{ref:note1}. As will become clear in the following, pumping currents are related to the coefficients in Eq.~\eqref{eq:Smatrix} through the spin-mixing conductance $G_{\mathrm{mix}}=G_r+iG_i$, where $G_r=\frac{e^2\mathcal{A}}{h\pi^2}\iint|\Delta S|^2dk_ydk_z$ and $G_i=\frac{e^2\mathcal{A}}{h\pi^2}\iint\mathrm{Im}[S_0^*\Delta S]dk_ydk_z$, where $k_y$ and $k_z$ are the transverse momenta and $\mathcal{A}$ the interface cross section. Similar to their counterparts in F, $G_r$ typically overwhelms $G_i$ by orders of magnitude.

By integrating over the Fermi surface, we obtain $G_r=G_r(\lambda,\delta)$ and plot it in the upper panel of Fig.~\ref{Fig:DeltaS}, where $G_r$ reaches the maximum at $\lambda=0.86$ and $\delta=0.5$. To elucidate how spin scattering is affected by the staggered field, we also calculate $G_r$ for an uncompensated interface as a representative for N$|$F and plot the result in the lower panel of Fig.~\ref{Fig:DeltaS}. Clearly, the two cases are similar in magnitude~\cite{ref:note2}, implying that spin transfer on a compensated N$|$AF interface is as efficient as that on N$|$F for the case of insulating magnets. With the current insight of AF dynamics and the reciprocity between spin pumping and STT discussed below, this feature is consistent with the expectations in Ref.~\cite{ref:Bauer} of ``no difference for the spin absorbed by a fully ordered interface with a large net magnetic moment or a compensated one.''

\textit{Spin pumping.}--- Although the AF resonance frequency reaches the THz region ($1\sim10$ meV), the motion of the staggered field remains adiabatic as evidenced by comparing ($\hbar$ times) the resonance frequency with two characteristic energy scales: (i) the Fermi energy in N is a few eV; (ii) the exchange coupling between conduction electron spins and magnetic moments can be as large as eV. As a result, the spin eigenstates and the scattering matrix Eq.~\eqref{eq:Smatrix} adiabatically adapt to the instantaneous configuration of AF. Regarding the staggered field $\bm{n}$ and magnetization $\bm{m}$ as two independent adiabatic parameters~\cite{ref:Brouwer}, we obtain the pumped spin current with the scattering matrix $S$ in Eq.~\eqref{eq:Smatrix}:
\begin{align}
  \frac{e}{\hbar}\bm{I}_s=G_r(\bm{n}\times\dot{\bm{n}}+\bm{m}\times\dot{\bm{m}})-G_i\dot{\bm{m}}, \label{eq:Is}
\end{align}
where $\bm{I}_s$ is measured in units of an electrical current. Since $\bm{n}=(\bm{m}_1-\bm{m}_2)/2$ and $\bm{m}=(\bm{m}_1+\bm{m}_2)/2$, Eq.~\eqref{eq:Is} can indeed be interpreted as arising from a coherent sum of two independent F spin pumping contributions by $\bm{m}_1$ and $\bm{m}_2$, which justifies the naive result envisioned at the beginning. However, the spin-mixing conductance $G_r$ and $G_i$ are \textit{different} from those of F due to the mixing of scattering channels from different sublattices. Moreover, AF dynamics is much faster than F thus a stronger spin pumping is expected from AF.

By taking a time average of Eq.~\eqref{eq:Is} over one period of oscillation, only the first two terms survive and contribute to the dc component of spin current $I_{s}^{dc}$. Despite that $|\bm{m}|\ll|\bm{n}|$, the contribution of $\bm{m}\times\dot{\bm{m}}$ to $I_{s}^{dc}$ can be comparable to that of $\bm{n}\times\dot{\bm{n}}$. This is because $I_{s}^{dc}$ is proportional to $\theta^2$ ($\theta$ labels the cone angle of precession) and the cone angle associated with the staggered field is much smaller than the one associated with the magnetization, $\theta_n\approx0$ but $\theta_m\approx\pi/2$, as shown in Fig.~\ref{Fig:Modes}.

\begin{figure}
   \centering
   \includegraphics[width=0.95\linewidth]{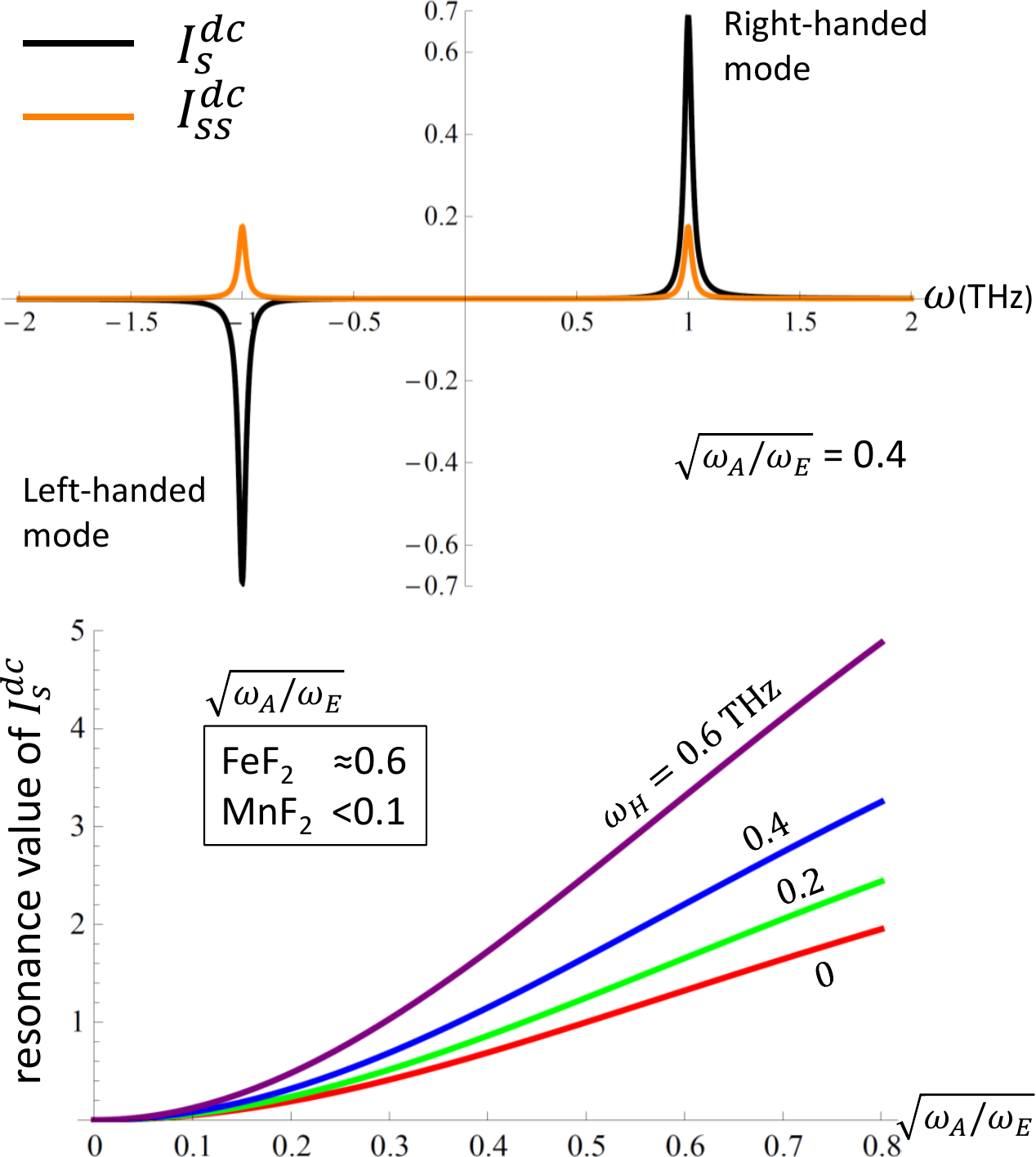}
   \caption{(Color online) Upper panel: dc components of spin and staggered spin currents as functions of $\omega$ in units of $\frac{\hbar}eG_r(\gamma h_{\perp})^2\cdot$ns. Parameters: $\omega_H\!=\!0$, $\omega_R\!=\!1$THz, $\sqrt{\omega_A/\omega_E}\!=\!0.4$, and Gilbert damping $\alpha=0.01$. Lower panel: for fixed microwave power, the resonance value of $I_{s}^{dc}$ (in the same unit as above) increases with increasing $\sqrt{\omega_A/\omega_E}$; it is also improvable by increasing $\omega_H$ ($-\omega_H$) when the right-handed (left-handed) mode is excited.}{\label{Fig:Absorption}}
\end{figure}

Consider now the AF motion generated by a microwave with oscillating magnetic field $\bm{h}_{\perp}$ perpendicular to the easy axis. If the microwave is circularly polarized, only the mode with matching polarization depicted in Fig.~\ref{Fig:Modes} is driven into resonance at certain frequency. When the magnetic field vanishes, $I_{s}^{dc}$ is an odd function of $\omega$ and is plotted in the upper panel of Fig.~\ref{Fig:Absorption}, where the peak (dip) for positive (negative) $\omega$ corresponds to the resonance of the right-handed (left-handed) mode. Hence, an important consequence is implied: the direction of the dc spin current is linked to the circular polarization of the microwave.

Since the sublattice degree of freedom is involved in the AF dynamics, we can also derive a staggered spin pumping. A staggered spin current represents the imbalance between the spin current carried by the two sublattices. It has three components $\bm{I}_{ss}^{(1)}$, $\bm{I}_{ss}^{(2)}$, $\bm{I}_{ss}^{(3)}$ associated with three pseudo-spin Pauli matrices. In a similar manner as spin pumping, we find that
\begin{align}
  \frac{e}{\hbar}\bm{I}_{ss}^{(3)}=G_r(\bm{n}\times\dot{\bm{m}}+\bm{m}\times\dot{\bm{n}})-G_i\dot{\bm{n}}, \label{eq:sIs}
\end{align}
and $\frac{e}{\hbar}\bm{I}_{ss}^{(1)}=-\mathrm{Im}[G_w]\dot{\bm{m}}$ and $\frac{e}{\hbar}\bm{I}_{ss}^{(2)}=-\mathrm{Re}[G_w]\dot{\bm{n}}$, where $G_w=\frac{e^2\mathcal{A}}{h\pi^2}\iint S_w^*\Delta Sdk_ydk_z$ results from inter-sublattice scattering that is unique to AF. When we take the time average, $\bm{I}_{ss}^{(1)}$ and $\bm{I}_{ss}^{(2)}$ drop out, only $\bm{I}_{ss}^{(3)}$ survives. This time, the dc component $I_{ss}^{dc}$ is an even function of $\omega$ in the absence of static magnetic field, which is plotted in Fig.~\ref{Fig:Absorption} (upper panel). We emphasize that elastic scattering in the normal metal will destroy any staggered spin accumulation, which decays on the time scale of $\hbar/t$. Therefore, the staggered spin current can only be defined within a distance of the mean free path away from the interface.

\textit{Detections.}--- When a spin current is injected into a heavy metal with strong spin-orbit coupling, it will be converted into a measurable transverse voltage via the inverse spin Hall effect~\cite{ref:ISHEHoffmann,ref:ISHEHillebrands,ref:ISHEAndo}. This effect has been widely used in the detection of spin pumping by F resonance, and we expect to verify our prediction with the same technique. However, in a recent experiment using Pt$|$MnF$_2$~\cite{ref:Ross}, no clear signal is found at a similar level of microwave power as in a conventional Pt$|$YIG. To explain this null observation, we resort to the efficiency of the microwave absorption at resonance point, which is proportional to $\sqrt{\omega_A/\omega_E}$ in an AF, whereas no such factor exists in a F. To see it more explicitly, we plot in Fig.~\ref{Fig:Absorption} (lower panel) the resonance value of $I_{s}^{dc}$ versus $\sqrt{\omega_A/\omega_E}$. In MnF$_2$~\cite{ref:MnF2}, $\sqrt{\omega_A/\omega_E}$ is only a few percent, which we believe is responsible for the suppression of the signals. Fortunately, there are better candidates, \textit{e.g.}, FeF$_2$ has the same crystal and magnetic structures as MnF$_2$, but the ratio $\sqrt{\omega_A/\omega_E}\approx0.6$ is extraordinarily large~\cite{ref:FeF2}. Thus, we expect a sizable microwave-driven spin pumping using Pt$|$FeF$_2$ heterostructure.

Small grains are unavoidable in large area N$|$AF interfaces since the typical grain size is below $\mu$m~\cite{ref:Ross}. As the optimal microwave absorption occurs only when the local easy axis is perpendicular to the oscillating magnetic field, the non-collinearity of the anisotropy fields of individual grains will somewhat reduce the net spin pumping upon averaging over the entire interface. However, progress in fabrication of N$|$AF heterostructures and reduced cross sections should lead to improved surface quality with less disorder in the form of grains.

The microwave absorption can also be enhanced by reducing the resonance frequency with a strong magnetic field, as illustrated by the lower panel of Fig.~\ref{Fig:Absorption}. But this brings about a challenge that it is hard to take full advantage of the high frequency (THz) and the high efficiency simultaneously.

\textit{Spin-transfer torques.}--- The reciprocal effect of spin pumping is STT, which describes the backaction that a spin current exerts on the AF. In linear response, an AF is driven by two thermodynamic forces $\bm{f}_n=-\delta F/\delta\bm{n}$ and $\bm{f}_m=-\delta F/\delta\bm{m}$ (energy dimension), where $F=(\hbar/2)\int d\mathcal{V}[\omega_0\bm{m}^2/a^3+\omega_n\sum_{i=x,y,z}(\partial_i\bm{n})^2/a-\omega_H\bm{H}\cdot\bm{m}/(Ha^3)]$ is the rree energy~\cite{ref:Landau}. Here we have scaled each term by the frequency in order to be consistent with our previous discussions; $\omega_0$ and $\omega_n$ are the homogeneous and inhomogeneous exchange frequencies, respectively. It can be easily shown that $\omega_0=\omega
_A+2\omega_E$. Enforced by $\bm{m}\cdot\bm{n}=0$ and $|\bm{n}|^2\approx1$, the symmetry allowed dynamics are: $\hbar\dot{\bm{n}}=(a^3/\mathcal{V})\bm{f}_m\times\bm{n}$ and $\hbar\dot{\bm{m}}=(a^3/\mathcal{V})[\bm{f}_n\times\bm{n}+\bm{f}_m\times\bm{m}]$~\cite{ref:Arne}, where $\mathcal{V}$ is the system volume. Inserting them into Eq.~\eqref{eq:Is} gives the response of the spin current to $\bm{f}_n$ and $\bm{f}_m$. Invoking the Onsager reciprocity relation~\cite{ref:Maekawabook}, we derive the response of $\bm{n}$ and $\bm{m}$ to a given spin voltage $\bm{V}_s$ in the normal metal which are identified as two STT terms $\bm{\tau}_n$ and $\bm{\tau}_m$. To linear order in $\bm{m}$, we obtain (frequency dimension)
\begin{subequations}
\begin{align}
 &\bm{\tau}_n=-\frac{a^3}{e\mathcal{V}}[G_r\bm{n}\times(\bm{m}\times\bm{V}_s)-G_i \bm{n}\times\bm{V}_s], \label{eq:taum}\\
 &\bm{\tau}_m=-\frac{a^3}{e\mathcal{V}}G_r\bm{n}\times(\bm{n}\times\bm{V}_s), \label{eq:taun}
\end{align}
\end{subequations}
which are consistent with the proposed phenomenological model~\cite{ref:Gomonay} that treats STTs on the two sublattices as completely independent.

In solving the AF dynamics, it is instructive to eliminate $\bm{m}$ and derive a closed equation of motion in terms of $\bm{n}$ alone~\cite{ref:Duine,ref:Arne,ref:Ran,ref:Haldane,ref:Ivanov}. Truncating to linear order in $\bm{V}_s$, $\bm{m}$, and $\dot{\bm{n}}$, we obtain the effective dynamics
\begin{align}
 \bm{n}\times(\ddot{\bm{n}}+\alpha\omega_0\dot{\bm{n}}+\omega_R^2\bm{n}_{\perp}) = \frac{\omega_0a^3G_r}{e\mathcal{V}}\bm{n}\times(\bm{n}\times\bm{V}_s), \label{eq:STT}
\end{align}
where $\alpha$ is the Gilbert damping constant, and $\bm{n}_{\perp}$ are perpendicular components of $\bm{n}$ with respect to the easy axis. Since the STT only acts on the interface and we consider a thin AF film, we have disregarded a possible nonuniform motion of $\bm{n}$; otherwise a term $\omega_0\omega_na^2\bm{n}\times\nabla^2\bm{n}$ should be included in Eq.~\eqref{eq:STT}. For thick metallic AF where electrons propagate into the bulk, Eq.~\eqref{eq:STT} should be replaced by its bulk counterpart~\cite{ref:Arne,ref:Ran}.

As an example, we consider the uniform AF dynamics driven by STT. Assume $\bm{V}_s$ is collinear with the easy axis, we solve the spectrum by virtue of Eq.~\eqref{eq:STT}: $\omega/\omega_0=\frac12[-i\alpha\pm\sqrt{-\alpha^2+4\omega_A/\omega_0+4ia^3G_rV_s/(e\mathcal{V}\omega_0)}]$. For small $V_s$, $\omega$ has a negative imaginary part so that any perturbed motion will decay exponentially in time and the system is stable. However, a sufficiently large $V_s$ will flip the sign of $\mathrm{Im}[\omega]$, which makes the system unstable and marks the onset of uniform AF excitation. By setting $\mathrm{Im}[\omega]=0$, we obtain the threshold spin voltage
\begin{align}
 V_s^{\mathrm{th}}=\pm\frac{e\mathcal{V}\alpha\omega_{_{\!R}}}{a^3G_r}, \label{eq:threshold}
\end{align}
where $+$($-$) corresponds to the excitation of the right-handed (left-handed) mode. The chirality selection by the sign of the spin voltage is just consistent with the direction control of spin pumping by the microwave polarization. Since $G_r$ scales linearly with the interface area, $V_s^{\mathrm{th}}$ scales linearly with the thickness of the AF layer.

In real experiments, a challenge arises from the large $\omega_{R}$, but we can still get reasonable $V_s^{\mathrm{th}}$ by reducing the layer thickness. For MnF$_2$ and FeF$_2$ of few nm thick, the threshold spin voltage is estimated to be 10-100~$\mu$V. The STT-driven AF dynamics suggests the feasibility of building a spin-torque nano-oscillator using AF, which generates a THz signal from a dc input without the need of static magnetic field.

We are grateful for insightful discussions with A. MacDonald, Y. You, E. Wahlstr\"{o}m, V. Flovik, M. Tsoi, H. Chen, Z. Qiu, and T. Ono. The work is supported by DOE-DMSE (No. DE-FG03- 02ER45958) and the Welch Foundation (No. F-1255). J. X. is supported by the special funds for the Major State Basic Research Project of China (Grant No. 2014CB921600).


\begin{thebibliography}{20}
   \bibitem{ref:Rasing} A. V. Kimel, A. Kirilyuk, P. A. Usachev, R. V. Pisarev, A. M. Balbashov, and Th. Rasing, Nature (London) \textbf{435}, 655 (2005); Nat. Phys. \textbf{5}, 727 (2009).
   \bibitem{ref:NiOswitch} T. Satoh \textit{et al.}, Phys. Rev. Lett. \textbf{105}, 077402 (2010).
   \bibitem{ref:Nowak} S. Wienholdt, D. Hinzke, and U. Nowak, Phys. Rev. Lett. \textbf{108}, 247207 (2012).
   \bibitem{ref:TAMR1} B. G. Park \textit{et al.}, Nat. Mater. \textbf{10}, 347 (2011); X. Marti \textit{et al.}, Phys. Rev. Lett. \textbf{108}, 017201 (2012).
   \bibitem{ref:TAMR2} Y. Y. Wang, C. Song, B. Cui, G. Y. Wang, F. Zeng, and F. Pan, Phys. Rev. Lett. \textbf{109}, 137201 (2012).
   \bibitem{ref:Tsoi} Z. Wei \emph{et al.}, Phys. Rev. Lett. \textbf{98}, 116603 (2007).
   \bibitem{ref:Urazhdin}S. Urazhdin and N. Anthony, Phys. Rev. Lett. \textbf{99}, 046602 (2007).
   \bibitem{ref:MacD} P. M. Haney and A. H. MacDonald, Phys. Rev. Lett. \textbf{100}, 196801 (2008); A. S. N\'{u}\~{n}ez, R. A. Duine, P. Haney, and A. H. MacDonald, Phys. Rev. B \textbf{73}, 214426 (2006).
   \bibitem{ref:Xia} Y. Xu, S. Wang, and K. Xia, Phys. Rev. Lett. \textbf{100}, 226602 (2008).
   \bibitem{ref:Duine} A. C. Swaving and R. A. Duine, Phys. Rev. B \textbf{83}, 054428 (2011); J. Phys.:Condens. Matter \textbf{24}, 024223 (2012).
   \bibitem{ref:Arne} K. M. D. Hals, Y. Tserkovnyak, A. Brataas, Phys. Rev. Lett. \textbf{106}, 107206 (2011); E. G. Tveten, A. Qaiumzadeh, O. A. Tretiakov, and A. Brataas, Phys. Rev. Lett. \textbf{110}, 127208 (2013).
   \bibitem{ref:Ran} R. Cheng and Q. Niu, Phys. Rev. B \textbf{86}, 245118 (2012); \textbf{89}, 081105(R) (2014).
   \bibitem{ref:Jacob} J. Linder, Phys. Rev. B \textbf{84}, 094404 (2011).
   \bibitem{ref:Gomonay} H. V. Gomonay and V. M. Loktev, Low Temp. Phys. \textbf{34}, 198 (2008); \textbf{40}, 17 (2014); Phys. Rev. B \textbf{81}, 144427 (2010).
   \bibitem{ref:Manchon} Hamed Ben Mohamed Saidaoui, A. Manchon, and X. Waintal, Phys. Rev. B \textbf{89}, 174430 (2014).  
   \bibitem{ref:STTReview1} D. C. Ralph and M. D. Stiles, J. Magn. Magn. Mater. \textbf{320}, 1190 (2008).
   \bibitem{ref:STTReview2} A. Brataas, A. D. Kent, and H. Ohno, Nat. Mater. \textbf{11}, 372 (2012).
   \bibitem{ref:spinpumping} Y. Tserkovnyak, A. Brataas, G. E. W.  Bauer, Phys. Rev. Lett \textbf{88}, 117601 (2002); Phys. Mod. Phys \textbf{77}, 1375 (2005).
   \bibitem{ref:Maekawabook} A. Brataas, Y. Tserkovnyak, G. E. W. Bauer, and P. J. Kelly, Chapter 8, \textit{Spin Current}, edited by S. Maekawa, S. Valenzuela, E. Saitoh, and T. Kimura (Oxford Univ. Press, New York, 2012).
   \bibitem{ref:Kittel} F. Keffer and C. Kittel, Phys. Rev. \textbf{85}, 329 (1952).
   \bibitem{ref:MnF2} F. M. Johnson and A. H. Nethercot, Jr., Phys. Rev. \textbf{114}, 705 (1959); M. Hagiwara, K. Katsumata, I. Yamada, and H. Suzuki, J. Phys. Condens. Matter \textbf{8}, 7349 (1996).
   \bibitem{ref:FeF2} R. C. Ohlmann and M. Tinkham, Phys. Rev. \textbf{123}, 425 (1961); R. C. Ohlmann, Ph.D. thesis, UC Berkeley, 1960.
   \bibitem{ref:note1} The term $\hat{\tau}_2[(\bm{n}\times\bm{m})\cdot\bm{\sigma}]$ is also allowed by symmetry. But its coefficient is much smaller than $\Delta S$. Meanwhile, it does not contribute to the mixing conductance upon integration over the Fermi surface.
   \bibitem{ref:note2} Within the tight-binding model, a bipartite AF is always insulating at half filling for finite $J$, regardless of $t_m$. But comparing to $t$ in N, $t_m$ is in general much smaller, thus $\delta=t_m/t$ is customarily taken to be close to 0. For $\delta\rightarrow0$, the maxima of $G_r$ appear at $\lambda=1$ for both compensated and uncompensated interfaces.  
   \bibitem{ref:Bauer}  X. Jia, K. Liu, K. Xia, and G. E. W. Bauer, Europhys. Lett. \textbf{96}, 17005 (2011).   
   \bibitem{ref:Brouwer} P. W. Brouwer, Phys. Rev. B \textbf{58}, R10135 (1998).
   \bibitem{ref:ISHEHoffmann} O. Mosendz, J. E. Pearson, F. Y. Fradin, G. E. W. Bauer, S. D. Bader, and A. Hoffmann, Phys. Rev. Lett. \textbf{104}, 046601 (2010).
   \bibitem{ref:ISHEHillebrands} C. W. Sandweg \textit{et al.}, Phys. Rev. Lett. \textbf{106}, 216601 (2011).
   \bibitem{ref:ISHEAndo} K. Ando \textit{et al.}, J. Appl. Phys. \textbf{109}, 103913 (2011).   
   \bibitem{ref:Ross} M. P. Ross, Ph. D. thesis, \textit{Spin Dynamics in an Antiferromagnet}, Technische Universitat Munchen, 2013.   
   \bibitem{ref:Landau} E. M. Lifshitz and L. P. Pitaevskii, \textit{Statistical Physics}, Course of Theoretical Physics Vol. 9 (Pergamon, Oxford, 1980), Part 2.
   \bibitem{ref:Haldane} F. D. M. Haldane, Phys. Rev. Lett. \textbf{50}, 1153 (1983); \textbf{61}, 1029 (1988).
   \bibitem{ref:Ivanov} I. V. Bar'yakhtar and B. A. Ivanov, JETP \textbf{58}, 190 (1983); Sov. J. Low Temp. Phys. \textbf{5}, 361 (1979).
\end{thebibliography}
\end{document}